\documentstyle{article}

\begin{document}

\vspace{0.2cm}

{\bf Comments on the paper by E. Gjerlow and O. Elgaroy ``Are all 
modes created equal ? An analysis of the WMAP 5- and 7-year data} 
\centerline{{\bf without inflationary prejudice"}}

\vspace{0.5cm}
\centerline{L. P. Grishchuk}
\centerline{School of Physics and Astronomy, Cardiff University,
Cardiff, CF24 3AA, United Kingdom\\}
\centerline{Sternberg Astronomical Institute, Moscow State University, Moscow,
119899, Russia}

\date{\today}

\vspace{0.8cm}
\begin{abstract}
The amount and characteristics of quantum-mechanically generated relic 
gravitational waves and primordial density perturbations is a subject of 
great theoretical and observational importance. Unfortunately, this subject 
is deeply contaminated by inflationary misunderstandings and incorrect 
``standard inflationary results". This note presents comments on a 
particular paper, arXiv:1008.4471v1. However, the comments may have a more 
general significance and may be of interest to other researchers working in 
this area of science.
\end{abstract}

\vspace{0.8cm}

In order of appearance in the text, my comments are as follows:

\vspace{0.5cm}

1. Zhao et. al. never claimed a detection, significant or semi-significant.
They speak about indications, and they quantify their claims in terms of 
confidence intervals. It is unfair to distort their position to such an extent 
as if they claimed detection, and then formulate the criticism as ``no such 
mode is present at a detectable level" (abstract), ``we first looked at the 
claims made in [5] and [6] concerning the detection" (conclusions), etc. 

2. Gjerlow and Elgaroy ``disagree with their claim that when looking for 
gravitational waves, one should only include the multipoles that are affected 
by them" (p.2). I think Zhao et. al. are right. If we knew in advance that
$A_s$ and $n_s$ at small scales are exactly the same as at large scales, then
this information could give better constraints on $R$. But the whole point is 
that $n_s$ has not to be the same, and Zhao et. al. provide specific evidence
from the actual data (partially seen also in revised analysis of the 
authors) that $n_s$ is probably not the same. Gjerlow and Elgaroy should 
reconsider their statement. It is likely that their position will change 
from disagreement to agreement. 

3. It seems to me that the revised calculations by Gjerlow and Elgaroy 
qualitatively confirm the results of Zhao et. al., even if Gjerlow and Elgaroy 
choose to describe them as a disagreement. But the central point of their 
criticism, which is the extension of the argument to their own ``testing the 
hypothesis of an $\ell$-dependent $n_s$" (p. 5) appears to be wrong. 
Apparently, the authors do not understand that the ``step-like" index $n_s$
of Zhao et. al. means a continuous spectrum with two power-law intervals. 
Instead, Gjerlow and Elgaroy make calculations for a discontinuous spectrum 
with two different $n_s$ and the same $A_s$. This calculation has little or 
no value. A correct calculation must first be performed, and then it will be 
seen whether their critical conclusions survive or not. 

4. The authors should make it clear that their expectations of what they 
should see from the test in sec.VB (or will see from the corrected test), 
and what so far they found as ``have not been met, or 
only partially" (p. 6, left), are not the expectations of Zhao et. al.. What 
Gjerlow and Elgaroy present as the refutation of the proposal of Zhao et. al. 
looks like the refutation of correct expectations from any combined analysis 
of two adjacent data sets. Certainly, there is no reason to expect 
``higher values for $R$ when using a step-like 
spectrum" (p. 6, right), because the likelihood function for $R$ in the 
combined set of data becomes flat, and the maximum likelihood (ML) point 
cannot serve as a reliable criterion, it can happen to appear almost anywhere. 
And certainly one should not expect a ``significantly" better fit to data 
(p. 5, right), because all the indications of the changing $n_s$ are still 
weak; the $\chi^2$ will improve somewhat, but not ``significantly".

5. The authors stress many times that they made a ``closer scrutiny" of 
the claims (abstract), ``improved versions" of the analysis (p. 1), that they 
have used ``the exact likelihood functions from WMAP in contrast to the 
approximations made in those papers" (p. 7), and ``the official noise values 
instead of approximated values for 
the noise" (conclusions), etc. All these formulations hint at a much 
more reliable analysis, than that of Zhao et. al.  It seems to me, 
however, that in reality all these words mean only that Gjerlow 
and Elgaroy have used a larger black box, called CosmoMC. The likelihood 
function derived by Zhao et. al. is simplified in its treatment of noises, 
but it is transparent and it is based on the original Wishart distribution, 
and not on various Gaussian approximations to this distribution that are 
adopted in the WMAP and CosmoMC software. If the difference in results 
were the matter of better treatment of noises, the posterior distributions 
for $R$ would probably increase their spread without changing the ML values, 
but this is not what happened. As the authors discovered, ``the values for 
$R$ are consistently lower than those found in the analysis of 
Zhao et. al." (p. 5, left). Surely, Gjerlow and Elgaroy know the concrete 
answer to the question why the ML values of $R$ in their analysis turned out 
to be systematically almost twice lower than in the similar analysis 
of Zhao et.al.. If so, Gjerlow and Elgaroy must include their answer in the 
paper. If not, they should not present their results as more reliable, if 
they are only different for unclear reason. 
 
\vspace{0.5cm}

The second part of the paper describes the authors' theoretical thoughts.
The general goal seems to show that it is not only that there 
are no indications of gravitational waves in the current data, but that 
there should not be any indications at all, because Grishchuk is wrong and 
the ``standard result" is correct. I do not want to be rude, but Gjerlow and 
Elgaroy simply do not understand the problem which they try to offer their 
thoughts on. 

\vspace{0.5cm}

6. They start demonstrating their incompetence right from the defining eq.(6), 
where they think ``$Q$ becomes the amplitude of the perturbations" (p. 10, 
below eq.(6)). In reality, $Q$ has nothing to do with the amplitude of the
perturbations, it is a complex space-dependent eigen-function with absolute 
value equal to 1.

7. In sec.VIA they try to convince themselves and the reader that Grishchuk
has made a crucial mistake in his calculations. With the words ``since we 
have" they introduce eq.(14). This is the equation that was derived by 
Grishchuk and which was the final destination of the entire debate. This 
equation answers all the questions: the quantity $\zeta$ before the transition 
is equal to the quantity $\zeta$ after the transition. There is no any 
claimed by inflationists ``big amplification during reheating", which would 
make $\zeta$ many orders of magnitude larger than the gravitational-wave 
amplitude $h$. [This excess of the resulting $\zeta$ (``scalar" perturbations) 
over $h$ (``tensor" perturbations) by many orders of magnitude is precisely 
what the so-called standard inflationary result wants to dig out from 
somewhere.] Having in front of them the ``we-have-equation" (14), which 
answered all the questions, Gjerlow and Elgaroy nevertheless 
insist (together with the equally confused predecessors) that Grishchuk 
has made a huge mistake. They begin a forensic study of the 
question whether Grishchuk ``did implicitly use 
the assumption of a continuous $\mu$" (p.11, left, middle). 
The ``we-have-equation" (14) tells them that if it were true that 
Grishchuk used the assumption of a continuous $\mu$ then he would not be
able to derive the equality (14), because the quantity $\gamma$ jumps 
by many orders of magnitude at the transition point. Nevetheless, Gjerlow and 
Elgaroy, in absolute conflict with their own ``we-have-equation" (14) 
(derived by Grishchuk), insist that Grishchuk missed a huge ``amplifying 
factor which propagates through the rest of Grishchuk's derivation, again 
yielding the standard result" (p. 11, left, bottom). They do not 
understand that the ``we-have-equation" (14) is the full answer to the 
discussed problem, but they announce, in a quite unacceptable manner, that  
``Grishchuk's treatment does not hold up under closer scrutiny" (p.11, left). 
I can only suspect that Gjerlow and Elgaroy do not understand even the 
question, let alone the solution, which is being discussed.

8. Having proved that there is no any ``big amplification during reheating",
Grishchuk explains that the only way to arrive at the arbitrarily large 
resulting $\zeta$ (this is what inflationists want) is to postulate this 
arbitrarily large $\zeta$ from the very beginning, as the initial condition.
In the quantum version, - as the initial squeezed vacuum (multiparticle) 
state. Of course, there is absolutely no physical justification for such a 
choice. In sec.VIB, Gjerlow and Elgaroy (together with the equally confused 
predecessors) start new round of confusion around this absolutely clear 
situation. 

The authors return to $\zeta$. They say that 
``Grishchuk claims that this variable is equal to zero" (p.11, right). This 
is an unbelievable distortion, and probably deliberate. One must 
have no understanding of the subject at all to claim that the 
variable $\zeta$ is equal to zero. Grishchuk derives equation for the 
function $\zeta$ and shows that there is no any claimed conservation law 
for $\zeta$: the constant part of $\zeta$ independent of initial conditions 
must be equal to zero, not the variable itself is equal to zero. Then 
``Grishchuk seems to agree ...that $\zeta$ is both constant and 
nonzero" (p.11, right). Of course he does, 
because this is what he was always saying (if you understand what you are 
asking). Grishchuk shows that there is nothing special in evolution 
of $\zeta(\eta)$, it behaves in exactly the same manner as the gravitational 
wave function $h(\eta)$. Namely, if the background scale factor and the 
initial conditions are such that one of the two independent solutions can be 
neglected, then $\zeta(\eta)$ during this interval of time is approximately 
constant and nonzero. In particular, this is what is reflected in the 
``we-have-equation" (14).

Then Gjerlow and Elgaroy enter the quantum-mechanical discussion. It is 
quite embarassing to comment on this part of their paper, because from the 
occasional formulations such as ``quantum states that are used for 
quantization", ``new states that satisfy the ground state 
condition", ``the expectation value of this variable to be that 
of the ground state of an harmonic oscillator" (p. 11, right, middle) it 
becomes clear that they scarcely have a clue what they are talking about. 
They do not understand that eq.(16) is not a normalization of $\zeta$ 
(which should be determined by initial conditions or quantum state) but a 
convenient technical redefinition of $\zeta$ by way of absorbing the 
known constants. They do not understand that the reprimanding phrase 
``Grishchuk blamed the states, when he should have blamed the 
normalization of $\zeta$" (p. 11, right, middle) is a nonsense, because the 
states and normalization are the same thing in the discussed context. They 
claim that Grishchuk ``says that when moving from the treatment of 
gravitational waves to density perturbations, one should always do the 
replacement $a \rightarrow a \sqrt{\gamma_i}$" (p. 11, end - p.12, beginning). 
This is a completely unacceptable distortion, and possibly deliberate. One 
must be fully confused in the subject to think that  
this replacement should ALWAYS be done. Grishchuk shows that the coupling 
function in the equations obeys this rule. But the scale factor $a$ enters 
also the universal definition of the wavelength, $\lambda = 2 \pi a/n$, and 
it would be an unforgivable mistake to do this replacement here, because 
this would make the wavelengths of the compared gravitational wave and 
density perturbation differ by many orders of magnitude. The initial 
wavelength participates 
in the redefinition (16). Gjerlow and Elgaroy noticed the letter $a_0$ in 
eq.(16). Since they keep in mind this wrong idea (taken from somewhere, 
not from Grishchuk's papers) that one should ALWAYS replace $a$ as stated 
above, they demand that Grishchuk should make this replacement in eq.(16), 
thus generating a huge factor that they want to see. Since Grishchuk refuses 
to do this unforgivable error, they announce that ``the normalization 
of $\zeta$ in Grishchuk's paper seems inconsistent with what he himself 
says" (p. 11, right). 

I am sorry, I cannot call the criticism in sec.VIB by any other words than 
total nonsense and defamation.

9. Gjerlow and Elgaroy used the combination of words ``standard calculation" 
many times, from the abstract to the end of the paper. They pretend to be 
honestly interested in finding out what is going on in this dispute. If so, 
it should be so easy to do this. Simply present the ``standard calculation" 
and show what kind of a silly mistake Grishchuk proposes to do in the 
``standard calculation". However, this never happened. Together with confused 
predecessors, Gjerlow and Elgaroy strive to find a huge mistake in Grishchuk's 
calculation. This is not accidental. There simply does not exist any ``standard 
calculation". What does exist is a set of logical jumps, incorrect transfers 
of notions and results from one theoretical framework to another, unjustified 
physical assumptions, and so on. In the end, Gjerlow and Elgaroy declare that 
``Grishchuk's claims were firmly rebutted and that he has not responed 
adequately to these rebuttals". 

I do not think that Grishchuk should chase every accuser with a gun, 
trying to respond adequately to the alleged rebuttals. All the raised 
questions and ``rebuttals" were many times discussed and responded to in 
publications. The summarizing and the latest one is:  L.P.Grishchuk, 
in {\it General Relativity and John Archibald Wheeler}, Eds. I.Ciufolini 
and R.Matzner, (Springer, 2010, pp.151-199) [arXiv:0707.3319]. What is 
improper indeed, is the fact that Gjerlow and Elgaroy did not 
take the trouble of reading this material. They are certainly aware of 
its existence, because they give reference [28] in the list of references. 
But they prefer not to even mention this reference in the text itself of their 
paper. If they are really interested in the correct treatment of this 
problem, as well as in the adequate responses to the alleged 
rebuttals, Gjerlow and Elgaroy must study and properly quote this paper. 
I think their public defamation of theoretical work of Grishchuk is 
unacceptable. 

\vspace{0.5cm}

In my opinion, this paper is not publishable in its present form. The authors
should first address all the issues raised in the above comments 1-9.

\end{document}